\documentclass{mn2e}

\usepackage{graphicx}

\def \nusym{\nu_{\rm sym}}
\def \rmaxtwo{r^{\rm max}_2}
\def \dcut{\delta_{\rm cut}}
\def \chin{\chi^2/dof}

\def \zobs {\zeta} %{\zeta_{\rm obs}}
\def \psipa {\psi_{\rm PA}}

\def \nucr {\nu_{\rm crv}}

\def \phim {\phi_m}

\def \mref#1{(\ref{#1})}

\def \nts{\negthinspace}

\newcommand{\rlc}{R_{\rm lc}}

\def \rns{R_{\rm NS}}

\def \numin {\nu_{\rm min}}
\def \numax {\nu_{\rm max}}

\title[Bifurcated components in pulsar profiles]
{Asymmetry of bifurcated features in radio pulsar profiles}

\author[J.~Dyks, and B.~Rudak
]{J. Dyks %$^{1}$,
and B.~Rudak\\%$^{1}$\\
Nicolaus Copernicus Astronomical Center, Toru\'n, Poland\\
}
\begin{document}

%\date{Accepted 1988 December 15. Received 1988 December 14; in original form 1988 October 11}
\date{Accepted 2011 November 25. Received 2011 Novemver 24; in original form 2011 August 23}

%\pagerange{\pageref{firstpage}--\pageref{lastpage}} \pubyear{2002}

\maketitle

\label{firstpage}

\begin{abstract}
High-quality integrated radio profiles of some pulsars contain
bifurcated, highly symmetric emission components (BECs). 
They are observed when our
line of sight traverses through a split-fan shaped emission beam.   
It is shown that for oblique cuts through such a beam, the features appear
asymmetric at nearly all frequencies, except from a single `frequency of
symmetry' $\nusym$, at which both peaks in the BEC have the same height.
Around $\nusym$ the ratio of flux in the two peaks of a BEC evolves
in a way resembling the multifrequency behaviour of J1012$+$5307.
Because of the inherent asymmetry resulting from the oblique traverse of
sightline, each minimum in double notches can be modelled 
independently. Such a composed model reproduces the double notches
of B1929$+$10 if the fitted function is the microscopic beam of curvature
radiation in the orthogonal polarisation mode. 
These results confirm our view that 
%at least
some of the double components
in radio pulsar profiles directly reveal the microscopic nature of 
the emitted radiation beam as the microbeam of curvature radiation polarised
orthogonally to the trajectory of electrons.
\end{abstract}


\section{Introduction}

Double absorption 
features have so far been observed
in integrated radio profiles of 
B1929$+$10 (Rankin \& Rathnasree 1997),
J0437$-$4715 (Navarro et al.~1997)
and B0950$+$08 (McLaughlin \& Rankin 2004).
They are the peculiar `W'-shaped features observed in highly polarised
emission. The minima of the `W' approach each other 
with increasing frequency $\nu$.
They have a large depth of $20-50$\% (Perry \& Lyne 1985; Rankin \&
Rathnasree 1997).

Initial attempts tried to explain the features in terms of a double eclipse,
with the doubleness generated by special relativistic effects (Wright 2004;
Dyks, Fr\c ackowiak, S{\l}owikowska, et al.~2005). Dyks, Rudak \& Rankin
(2007, hereafter DRR07) proposed that the feature represents the shape of 
a microscopic beam of emitted coherent radiation. 
It has been suggested that this beam is also observable in pulsar profiles
as a double emission
feature, hereafter called a bifurcated emission component (BEC).
Examples of such components can be found in the pulse profiles of 
J0437$-$4715 and J1012$+$5307 (Dyks, Rudak \& Demorest 2010, hereafter DRD10).
However, the beam that had been proposed initially by DRR07 -- 
the hollow cone of emission caused by parallel acceleration -- 
could not reproduce the observed large depth of double notches ($20-50$\%).

DRD10 have shown that the drop of flux in the minima of double notches
by %nearly 50\% 
several tens percent
imposes constraints on the general structure of the emission beam.
The flux drop in the `W' must be caused by either a small non-emitting region
(wherein the coherent amplification fails),
or by some obscuring/intervening/eclipsing object.
For the hollow cone beam, the decrease of flux expected in such a model
is maximally a few percent (see figs.~2c, and 4 in DRR07).
This is because within the hollow-cone beam 
a considerable amount of radiation is emitted along
many non-parallel directions.
A single line of sight is then
capable of receiving considerable contributions from sideway emissions.
For this reason
a laterally extended, two- (or three-) dimensional emission region contains 
many places that backlit the notches and make them shallow.
Numerical simulations have shown that this ``depth problem"
cannot be naturally overcome through manipulations with geometry of the
non-emitting (or absorbing) region (see fig. 5 in DRR07 for sample
results).

In DRD10 it has been recognized that the depth approaching $\sim50$\% 
results from the fact that the emission is intrinsically
two-directional, or, that the beam is double-lobed. In such a case, at any
pulse longitude the observer
simultaneously detects radiation from two separate places 
located somewhere within the extended emission region. 
At the phase of minimum in the notches, 
only one of the two places is eclipsed (or is non-emitting), hence the 50\%
upper limit for the depth of notches.

\begin{figure}
   \includegraphics[width=0.49\textwidth]{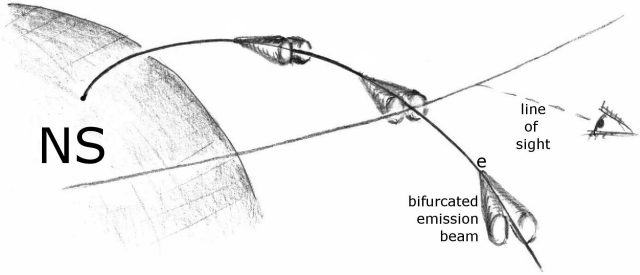}
       \caption{The principle of creating double features in radio pulse
profiles: the line of sight traverses through a split-fan emission pattern, 
produced by outflowing particles that emit a double-lobed beam. For simplicity,
the figure shows only a single trajectory of electrons flowing in a stream 
of finite width.}  
      \label{stream}
\end{figure}

When the double-lobed beam is carried along with the electron, a split-fan beam
is created, as shown in Fig.~\ref{stream}.
In this paper we show that when the line of sight cuts through the stream
at an oblique angle, a specific change of the BEC's shape is expected 
across the frequency spectrum (Section 2). 
The resulting asymmetry of the BEC allows for
a piece-wise fitting of double notches with a physical beam-shape model
(Section 3).

One important difference between 
the stream-cut origin of components and the conal view of profile 
formation is the following: the \emph{altitude} of detectable
emission corresponds to that part of the stream, which is nearly tangential
to the line of sight, whereas in the conal model, the altitude is fixed
by a preselected value of $\nu$ (the tangency condition 
determines the \emph{azimuth} of detectable region, but not the altitude).
In the case of the stream-cut scenario, when the line of sight is not in 
the plane of the radio-emitting
stream, no (or little)
radiation is detectable, irrespective of $\nu$. 
When the sightline crosses the stream plane, only a localized 
piece of the stream is detectable, namely, the part that happens to be
(nearly) tangent to the line of sight.
Thus, the `observable' altitude is 
frequency-independent, because the condition for tangency 
does not depend on the frequency of observation.
Consequently, if the components observed in pulse profiles
arise from stream cutting
(whether they appear double, or single because of convolution effects),
they will not exhibit the radius-to-frequency mapping (RFM).
That is, the components will remain at fixed, frequency-independent
locations in pulse profiles observed at different $\nu$.

Such a stream-cut origin of components allows us then to explain
the lack of RFM in millisecond pulsars (MSPs; Kramer et al.~1999). 
Stream-cut viewing geometry is more likely in MSPs, because 
the polar cap of MSPs is much larger than for normal pulsars.
Therefore, any transverse motion of limited magnitude, 
like the $\vec E \times \vec B$ 
drift, is less likely to create the hollow-cone emission pattern.
Moreover, magnetic field lines in the MSPs 
flare strongly away from the dipole axis, 
which facilitates cutting of sightline through streams. 

\begin{figure*}
   \includegraphics[width=0.88\textwidth]{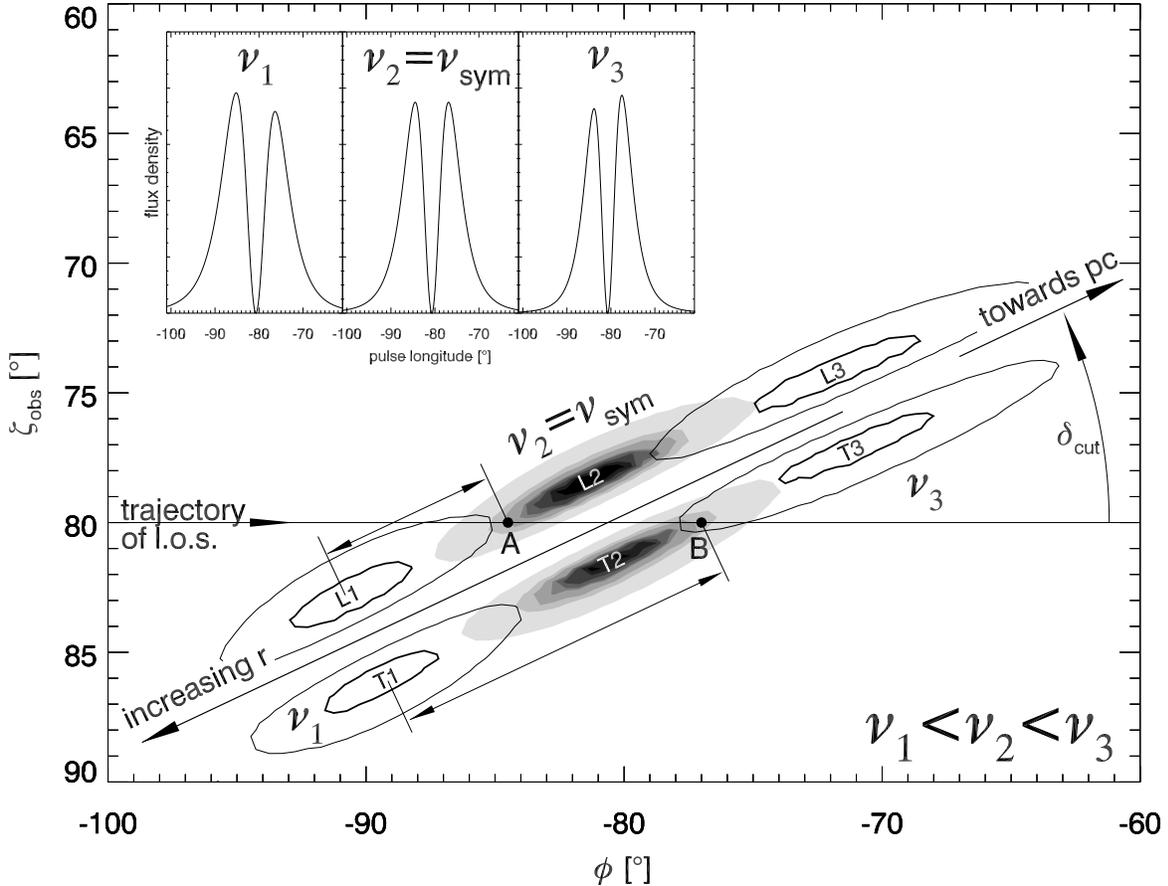}
       \caption{Origin of asymmetry inversion in bifurcated components of
pulsar profiles. The three pairs of contours present sky-patterns of flux
at three frequencies $\nu_1 < \nusym < \nu_3$. The patterns result from
emission of a stream extending quasi-diagonally towards the bottom-left
corner of the graph. Observer's line of sight crosses the stream along 
the horizontal line. For more details see Section \ref{oblique}.
       }
      \label{show3b}
\end{figure*}

\section{Oblique cut through split-fan beams}
\label{oblique}

The split-fan emission of a stream, when projected on the sky,
creates an elongated double track, such as 
the slanted pattern shown in
Fig.~\ref{show3b}.
% or the radially outspreading tracks of Fig ... .
%Fig.~\ref{show3b} 
There, the sky-projected emission is presented at three frequencies:
$\nu_1$ (two left contours: L1, and T1), $\nu_2$ 
(grey central contours: L2, T2), 
and $\nu_3$ (two right-hand side contours: L3, T3). 
For clarity of the figure,
the contours mark only the brightest (topmost) parts of the emission
pattern. There is actually considerable (and detectable) level of emission
extending all the way across the figure at each frequency.
Passage of sightline through the pattern, eg.~along the
horizontal line in Fig.~\ref{show3b},
results in essentially double shape of observed component, 
regardless of
the frequency of observation (see the triple inset).
Assuming some residual RFM, for decreasing $\nu$ the double track 
has a maximum at increasing distances from the polar cap, ie.
the location of the peak flux moves from right to left in Fig.~\ref{show3b}.
%Assuming that the radial extent of emission is sufficiently large,
For appropriate traverse of sightline, 
there exists a frequency (denoted $\nu_2$ in Fig.~\ref{show3b}, and
hereafter called the frequency of symmetry, $\nusym$),
at which both maxima in the BEC have the same peak flux.
Note that for non-orthogonal cuts ($\dcut \ne 90^\circ$)
through the split pattern, the line of sight does not
cross through the maxima of the $\nusym$-pattern (marked `L2' and `T2' in the
figure). Even at this frequency of symmetry the sightline misses
the leading maximum L2, moving on the `far side' of it. 
That is, the sightline passes through the stream 
at a larger radial distance $r_{\rm A}$ 
than that of the peak emission: $r^{\rm max}(\nu_2)
\equiv\rmaxtwo< r_{\rm A}$.
The trailing maximum of the $\nu_2$-pattern is passed-by at $r_{\rm B} < 
\rmaxtwo$,
ie.~below the radial distance of peak flux at this frequency.
Thus, \emph{even when peaks of an observed BEC have perfectly 
the same-height, they nevertheless correspond to different emission altitudes} 
(for $\dcut \ne 0$). For this reason, despite having equally high peaks at
$\nusym$, the observed
BEC will not be \emph{ideally} symmetric: the peaks can have somewhat 
different 
widths due to a difference in curvature radii (see below).

At a lower frequency $\nu_1 < \nusym$ the BEC will have the leading peak
brighter than the trailing one, because our line of sight passes closer to
the leading maximum in the $\nu_1$ pattern (L1). Using the markings of
Fig.~\ref{show3b}, at frequency $\nu_1 < \nusym$ we have $|\rm{L1\ A}| < 
|\rm {T1\ B}|$.
This holds (ie.~the leading peak is the brighter one at `subsymmetric' 
frequency)
if the polar cap (region of lower altitudes) is located 
on the right-hand side (trailing side) of the observed feature, 
as in Fig.~\ref{show3b}. That is, the leading peak is expected to dominate at 
$\nu < \nusym$ if the BEC is observed as a precursor;
for a `postcursor' position of the BEC, the opposite is expected: a brighter
trailing peak at $\nu < \nusym$.

Conversely, the trailing peak of a BEC becomes brighter at $\nu_3 > \nusym$,
because $|\rm{A\ L}3| > |\rm{B\ T}3|$, see Fig.~\ref{show3b}.
Thus, when the BEC is inspected at increasing frequencies, \emph{the dominating
(brighter) peak switches from the leading to trailing position at
$\nusym$} (see the inset of Fig.~\ref{show3b}). Observations of Kramer
et al.~1999 (fig.~5 therein) provide evidence of such behaviour in
J1012$+$5307. 
\emph{For the precursor location of the BEC, it is the leading peak which 
is expected to be brighter at low $\nu$}, again in consistency with 
the profile of J1012$+$5307.
 
\subsection{Asymmetry of the width at $\nusym$}

At the ``frequency of symmetry" the maxima of BEC have precisely equal
height, but the symmetry is approximate, because their widths are slightly
different. The reason for this is that the width-scale of the BEC
also depends on the curvature of trajectory of electrons flowing
in the stream (in addition to the dependence on $\dcut$). Low in the dipolar 
field, the curvature radius $\rho$ 
increases with altitude, as a result of which the emitted beam becomes 
narrower. The leading half of the BEC originates
from larger altitudes (see Fig.~\ref{show3b}) and should therefore be
narrower than the trailing part, for which the altitude and $\rho$ 
are smaller. Assuming the oblique cut and the normal RFM,
the bifurcated \emph{precursors} are 
then expected to have the leading wing narrower than the trailing one.
Weak traces of such asymmetry can be discerned in the BEC
of J1012$+$5307 (see fig.~4 in DRD10). 

This inference has been verified with a numerical code that follows
%direction-resolved 
beam-resolved
emission of electrons moving along narrow 
streams in rotating dipolar
 magnetosphere. In the numerical calculations 
%of Section 2.1,
illustrated in Figs.~3 and 4,
we assume that the B-field has
the geometry of the rotating retarded vacuum dipole as given by equations
A1 -- A3 in Dyks \& Harding (2004)\footnote{We take the opportunity here
to warn that fig.~1 in Dyks \& Harding (2004), that compares 
the retarded and static-shape dipole, was corrupted by errors in
plotting routine and should be dismissed. All formulae in that paper,
however, are perfectly free from errors.}. Thus, we do not take into account
the modifications of $\vec B$, that are expected for plasma-filled
magnetosphere (see eg.~Bai \& Spitkovsky 2010 for the force-free case).
The usual static-shape dipole is assumed only for the  
analytical calculations of Section 2.2.2 and Fig.~5.
In the calculations that have led to Figs.~3 and 4, 
the local beam of non-coherent curvature radiation is emitted,
% from an array of points distributed along the stream, 
with the realistic size appropriate for 
a preselected frequency $\nu$, and for
local $\rho$ (as measured in the 
inertial observer's frame IOF).
However, instead of using the bandwidth-integrated beam,
in Figs.~3 and 4 
we have used the `bolometric beam' which is integrated over 
the entire frequency spectrum. This is because: 1) the bolometric beam is
given by the simple analytical formula of eq.~8 in DRD10, and 2)
the shape of the bolometric beam happens to be similar 
to the observed shape of the BEC in J1012$+$5307, despite the latter is
bandwidth-limited only.
%therefore it should in principle be similar to the frequency-resolved beam).
In this way we have avoided some deeply-nested calculation loops 
that would have otherwise been required to perform bandwidth-limited
integration of the frequency-resolved beam, which cannot
be expressed in a simple analytical way. 
Such a realistic beam shape (ie.~one which corresponds to the observed
bandwidth) is only used in Section 3 to fit the double notches of B1929$+$10.
%, cf.~eqs.~ 21 and 22 in Dyks, Wright & Demorest 2010) 
The Lorentz factor $\gamma$ in the bolometric beam is set to a value 
that corresponds to the realistic opening angle of bandwidth-limited 
beam at the central frequency of $\nu = 820$ MHz, and which also corresponds
to the local IOF curvature radius of electron trajectory.
[Note: In the low-frequency limit, ie.~when the BEC's spectrum extends 
to much higher frequency than the observation frequency $\nu$,
the frequency-resolved beam of curvature radiation does not depend
on the Lorentz factor, so the observation frequency and the curvature of
trajectory fully define the opening angle.]
%The-low frequency limit
%is applicable because the observed spectrum of the BEC of J1012 extends to
%several GHz.
Thus, whereas the shape of the beam used for calculations
illustrated in Figs.~3 and 4 is approximate, its opening angle is a
realistic one. The aberration and propagation time effects (for flat 
space-time) are included
in these calculations.

Note that on the LS of polar cap,
the IOF radius of curvature increases with the radial distance $r$ 
at a slower rate than the standard dipolar radius of curvature
$\rho_{\rm dip}$. The latter is equal to:
\begin{equation}
\rho_{\rm dip} \simeq \frac{4}{3}\frac{\sqrt{r\rlc}}{s},
\label{rhodip}
\end{equation}
where $s=\sin\theta_m/\sin\theta_{lo}$ is the fieldline footprint parameter,
$\theta_m$ is the magnetic colatitude (measured from the dipole axis),
$\theta_{lo}$ is the magnetic colatitude of the last open field lines
at radial distance $r$, and 
$\rlc$ is the light
cylinder radius. The corresponding radius of curvature of electron 
trajectory in the IOF is:
\begin{equation}
\rho \simeq \rlc\nts\left( \frac{3}{4} \frac{s}{\sqrt{r/\rlc}} + 2\sin\alpha 
\right)^{\nts\nts-1} \nts\nts\nts
= \rlc \nts\left( \frac{\rlc}{\rho_{\rm dip}} + 2\sin\alpha 
\right)^{\nts\nts-1}
\nts\nts\nts\nts\nts\nts.
\label{rhoiof}
\end{equation}
This equation only refers to the leading side of polar tube and holds when
the viewing angle $\zobs$ is equal to the dipole
inclination ($\alpha \simeq \zobs$).
Comparison of eqs.~\mref{rhodip} and \mref{rhoiof} tells us that 
the increase of radius of curvature with altitude
is smaller for the electron
trajectory, than it is for the corresponding B-field line along 
which the electron moves in the corotating frame CF.
For the orthogonally-rotating MSP with $P=5.25$
s ($\rlc = 25\cdot10^6$ cm), the dipolar radius of curvature 
at the leading side of polar cap is
$\rho_{\rm dip} = 6.7\cdot10^6$ cm, whereas the radius of curvature 
of the corresponding
IOF trajectory is equal to $\rho = 4.3\cdot10^6$ cm.
In our numerical code, the IOF values of $\rho$ are used. Within a limited
part of magnetosphere they are given by eq.~\mref{rhoiof} 
(Dyks, Wright \& Demorest 2010; Thomas \& Gangadhara 2005).

\begin{figure}
   \includegraphics[width=0.48\textwidth]{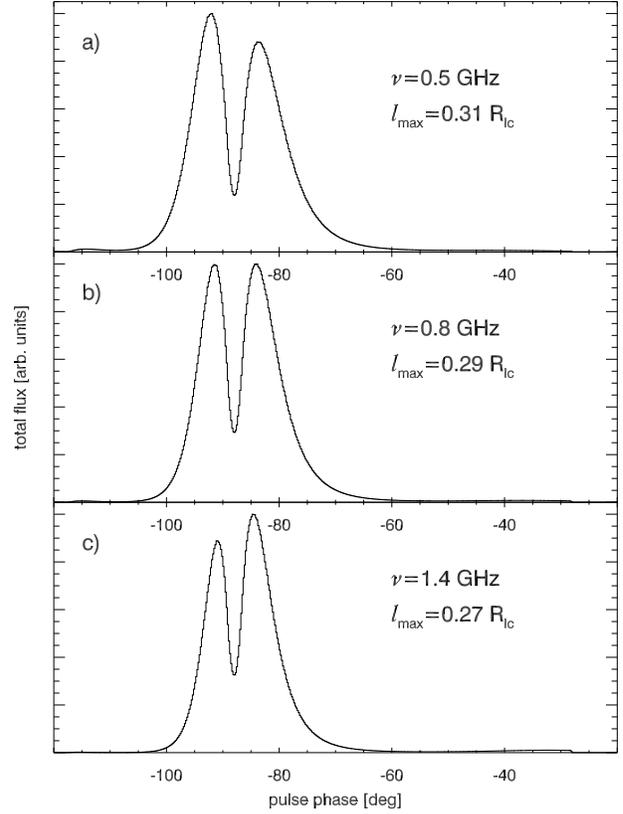}
       \caption{Frequency evolution of a BEC, calculated for $\alpha = \zeta
=85^\circ$, $\phi_m = 95^\circ$, $s = 1.2$, $P=5.26$ ms. The extent
of emissivity along the electron path was $\Delta l = 0.1\rlc$ and the 
RFM was $r \propto \nu^{-0.13}$.
Note the asymmetry inversion, and the increase of $\Delta$ for decreasing
$\nu$. This BEC is created by the viewing geometry represented 
by the lowest horizontal line in the next figure.}
      \label{r3}
\end{figure}

A sample result, presented in Fig.~\ref{r3}, has been obtained for an almost
orthogonal rotator, with dipole tilt of $\alpha = 85^\circ$,
$\zobs=85^\circ$, $P=5.25$ ms, $\nu =
0.5$, $0.8$, and $1.4$ GHz. The stream was infinitesimally thin
and constrained to a single B-field line with the footprint parameter 
$s = 1.2$ and the surface magnetic azimuth $\phi_m = 95^\circ$.
For the model of Fig.~\ref{show3b} to work,
the radial profile of emissivity may be represented by any
non-monotonic function of altitude, with a single maximum at some height.
A convenient function that we had arbitrarily selected 
was the Gauss function of the IOF length
of electron path (measured along their IOF trajectory). The emissivity
had a maximum at some `path-tracing' distance of $l_{\rm max}$ 
from the polar cap,
and the $1\sigma$ width of the emissivity profile was equal to $\Delta l =
0.1\rlc$.
A weak, residual RFM of $r \propto \nu^{-0.13}$ was assumed, 
ie.~the peak emissivity for the frequencies $\nu = 0.5$, $0.8$ and $1.4$
GHz occurred at distances of $0.31$, $0.29$, and $0.27\rlc$ from the polar
cap. 

Results of the simulations shown in
Fig.~\ref{r3} confirm the expectations based on Fig.~\ref{show3b}:
because of the RFM, the ratio of fluxes in the two peaks becomes inverted,
but the BEC stays fixed at the same pulse phase at all frequencies.
The separation of maxima in the BEC increases for decreasing frequencies
($\Delta_{\rm bfc} \propto \nu^{-1/3}$),
which is caused purely by intrinsic properties of the emitted microbeam;
it is not caused by the RFM.
The trailing half of the simulated BEC is wider than the leading one
(if both are measured at half maxima). This results from the fact that the
leading part of the BEC on average originates from a larger $r$
than the trailing part (see Fig.~\ref{show3b}).
In the simulation shown, the asymmetry of width amounts to $\sim4$\%, which is
noticeably larger than observed for J1012. A close inspection of fig.~8
in DRD10 reveals asymmetry of width of $\sim1$\% magnitude.
This may suggest that the BEC of J1012 originates from non-dipolar (more
circular) B-field lines, or that the cut angle $\dcut$ is larger than in
the simulation (or both).

The degree of asymmetry of peak flux at $\nu\ne \nusym$, depends in 
a degenerate way on several factors, such as $\Delta l$, $\dcut$, and the
strength of RFM. To measure the flux asymmetry, let us introduce
the ratio $R_F = F_l/F_t$, where $F_l$ and $F_t$ denote the peak flux
at the leading and trailing maximum in the BEC.
At frequencies far from $\nusym$ the ratio $R_F$ can evolve with $\nu$ 
or stay constant, depending on the radial profile of emissivity.
%at the selected $\nu$. 
This is because for a fixed viewing geometry
$R_F$ simply represents the ratio of emissivities at two fixed altitudes:
$R_F(\nu) = F_\nu(r_{\rm A})/F_\nu(r_{\rm B})$, where the indices `A' and `B' refer to
the points in Fig.~\ref{show3b}. 

For small values of $\Delta l$ or $\dcut$ the asymmetry $R_F$ strongly
depends on the viewing angle $\zobs$, which can vary as a result of
precession, as well as on $\alpha$. A change of $\zobs$ by just $1^\circ$
(from the value of $85^\circ$ shown in Fig.~\ref{r3}, to $\zobs = 84^\circ$) 
changes 
the flux ratio from $R_F \sim 1$ (Fig.~\ref{r3}b) to $R_F\sim 0.4$. With the
additional asymmetry in the width of the peaks, such components would
probably be not identified as `bifurcated' at all (they look just `double').

%HERE COMPARISON TO KRAMER ET AL DATA ON J1012 PLUS THE FIGURE

\subsection{Apparent magnification by geometric effects}

The double features can be interpreted as a direct consequence
of structured shape of microscopic emission beam.
Let us consider that part of the curvature radiation beam
which is polarised orthogonally to the plane of electron trajectory.
Let the frequency of observation be much smaller than the peak frequency 
$\nucr$ of standard curvature spectrum: 
$\nucr = 7\ {\rm GHz}\ \gamma^3/\rho\rm[cm]$, where
$\gamma$ is the electron Lorentz factor.
Radiation of this type has the intrinsic bifurcation angle of:
\begin{equation}
2\psi = 0.8^\circ (\rho_7\nu_9)^{-1/3},
\label{psi}
\end{equation}
where $\rho_7 = \rho/(10^7\ {\rm cm})$, and $\nu_9 = \nu/(1\ {\rm GHz})$.
For dipolar curvature radii $\rho_7 \sim 1$, the beam is up to 10 times 
narrower
than the size of features observed in B0950, B1929, and J1012. 
If the physical model is correct, 
it implies that either $\rho \sim 10^4-10^5$ cm, or that
the observed width has been strongly enlarged by geometrical magnification.

\subsubsection{Probability of small-angle cut through the stream}

\begin{figure}
   \includegraphics[width=0.48\textwidth]{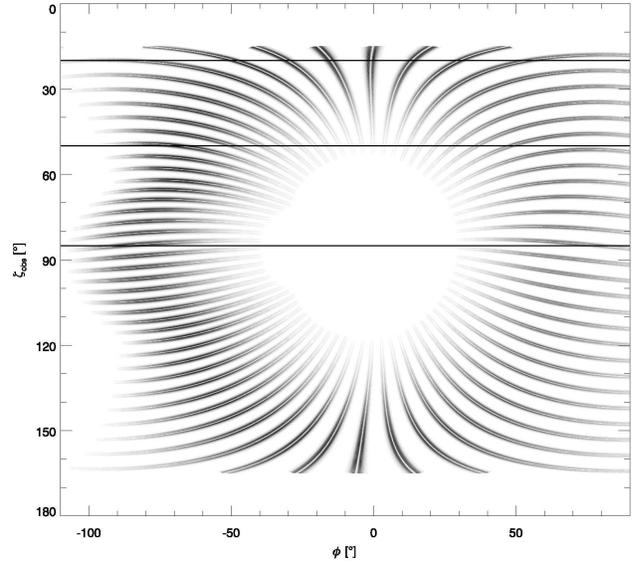}
       \caption{Sky-projected emission from 72 streams with $s = 1.2$
and $\phi_m = 0^\circ$, $5^\circ$, $10^\circ$, $15^\circ$, etc.
For all streams $l_{\rm max} = 0.25\rlc$ and $\Delta l = 0.1\rlc$. 
The other parameters are the same as in the previous figure.
Note the bifurcated nature of the emission stripes, and their 
tendency to assume horizontal
direction at large $|\phi|$, that results in a small cut angle $\dcut$.  
       }
      \label{skyd}
\end{figure}

To illustrate the effects of geometrical enlargement, we have calculated
a sky pattern of bifurcated emission from 72 streams that emerge from 
the surface
of neutron star at equal intervals of magnetic azimuth $\Delta\phim = 5^\circ$.
The result is shown in Fig.~\ref{skyd}, where the bifurcated nature of the
emission can be discerned as a low-flux fissure at the centre of each
emission stripe. We have selected B-field lines with the footprint $s=1.2$
to maximize the bifurcation angle (so it is more easily discernible 
in the figure)
as well as to avoid some overlapping and pile-up of emission due to the
caustic effects (which tend to occur for $s \la 1$).
The most vertical emission stripes correspond to $\phim = 190^\circ$ (the
top one), and $\phim = 350^\circ$ (bottom). 
The radial emissivity profile was the Gauss function with $l_{\rm max}
= 0.25\rlc$ and $\Delta l = 0.1\rlc$.
The three horizontals mark 
sightline
traverse at $\zobs = 20$, $50$, and $85^\circ$. The dipole tilt was $\alpha
=85^\circ$ so the radiation from the magnetic pole %at $r = \rns$
is located in Fig.~\ref{skyd}
at $\phi = -2\rns/\rlc = -4.5^\circ$, $\zobs = 85^\circ$. 

As can be seen in Fig.~\ref{skyd}, with increasing distance from the polar
cap the emission of streams
tends to assume horizontal direction and the stripes of emission
are cut at small angles by the
line of sight. Far away from the pole ($|\phi|\sim90^\circ$) this is true
for any viewing angle $\zobs$. For $\zobs \simeq \alpha$, the cut
angle is small within the entire phase range shown in Fig.~\ref{skyd},
except from the close vicinity of $\phi=0$. 
%for any pulse phase $\phi$. 
For small $\zobs \sim20^\circ$
the observed bifurcation is large for any phase, 
because of the classical magnification by the `not a great circle' effect.
Thus, for the case shown in Fig.~\ref{skyd}, it is only for moderate
viewing angles $\zobs \sim 30-60^\circ$, and for a limited range of phase
$|\phi|<70^\circ$, where the BECs are not strongly magnified.
Moreover, if we assume that easily-detectable pulsars are viewed
at a small impact angle ($\zobs \simeq \alpha$), then the strong enlargement
should be expected for a wide phase interval. This is because
in this case the cut angle $\dcut$ is noticeably large
only for limited phase range around $\phi \sim 0$, and $\phi \sim180^\circ$
(see the dashed line for $\zobs=75^\circ$ in Fig.~\ref{analbeam} further below).
%where the main pulse is likely 
%to complicate identification of any BEC.

We conclude that there is a fairly large parameter space in which the observed
width of BECs can be considerably increased by geometric effects of either
1) the small cut angle $\dcut$, or 2) the small viewing angle $\zobs$.
The presence of the first factor considerably increases the size of the
space, in comparison to the case of conal beam, for which
only the second type of enlargement was possible.

\subsubsection{Peak separation as a function of viewing geometry}

\begin{figure}
   \includegraphics[width=0.48\textwidth]{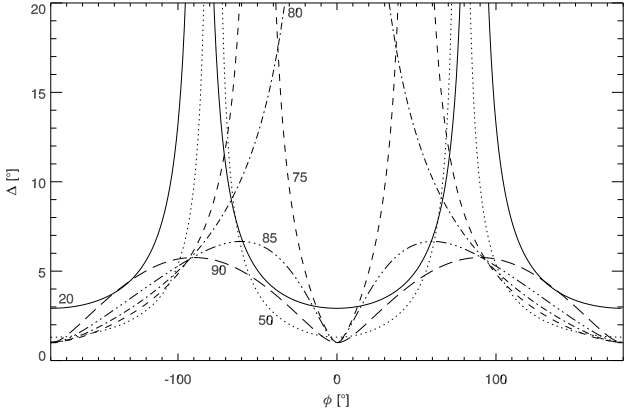}
       \caption{Observed separation of peaks in a double feature produced by 
the sightline's traverse through a split-fan emission beam
with bifurcation of $2\psi= 1^\circ$. Curves of different types 
are calculated from
eq.~\mref{delan} with $\alpha=80^\circ$, for six values of 
$\zobs = 20^\circ$ (solid), $50^\circ$
(dotted), $75^\circ$ (dashed), $80^\circ$ (dot-dashed), $85^\circ$
(three-dot-dashed), and $90^\circ$ (long-dashed).
Note the obvious possibility of observing $\Delta > 5^\circ$ within $|\phi| <
120^\circ$, despite $\psi= 0.5^\circ$.
       }
      \label{analbeam}
\end{figure}

If the aberration and retardation are neglected, the apparent separation
$\Delta$ of maxima in a BEC can be derived by application of spherical
trigonometry to the standard dipolar magnetic field.
The result is:
\begin{equation}
\cos{\left(\frac{\Delta}{2}\right)} = \frac{\cos\kappa - \cos^2\zobs}{\sin^2\zobs},
\label{delan}
\end{equation}
where
\begin{equation}
\sin{\kappa} = \frac{\sin{\psi}}{\sin\dcut}.
\label{kappa}
\end{equation}
The bifurcation angle $2\psi$ is given by eq.~\mref{psi}, whereas the cut
angle $\dcut$ is equal to:
\begin{equation}
\dcut = 90^\circ - \psipa,
\label{dcut}
\end{equation}
where $\psipa$ is the usual polarisation angle between the sky-projected 
B-field and the sky-projected rotation axis:
\begin{equation}
\tan\psipa = \frac{\sin\alpha\sin(\phi- \phi_0)}
{\cos(\phi-\phi_0)\cos\zobs\sin\alpha - \cos\alpha\sin\zobs}
\label{psipa}
\end{equation}
(Radhakrishnan \& Cooke 1969).
Equation \mref{delan} is the split-fan analogue of the well-known
equation for opening angle of conal beams.
When the bifurcation angle $2\psi$, and the observed separation $\Delta$
are small, 
eq.~\mref{delan}
 reduces to:
\begin{equation}
\Delta \simeq \frac{2\psi}{\sin\zobs\sin\dcut}
\label{delapp}
\end{equation}
as was guessed in DRD10.

With the special relativistic (and altitude-sensitive) effects ignored,
the cut angle $\dcut$ is uniquely determined for every pulse phase $\phi$
as soon as $\alpha$, $\zobs$ and $\phi_0$ are selected. Therefore,
eq.~\mref{delan} actually gives the separation $\Delta$ as a function 
of pulse phase $\phi$. Fig.~\ref{analbeam} presents examples
of such curves calculated for $\alpha=80^\circ$, $\psi=0.5^\circ$, and various
viewing angles $\zobs$. 
It can be seen that for $\zobs < \alpha$ our sightline's traverse becomes
instantaneously tangent to the projected B-field lines, which causes
$\Delta$ to increase infinitely. For small $\zobs \ll \alpha$ 
this happens near $|\phi|\sim90^\circ$ whereas for $\zobs$ approaching 
$\alpha$ this phase interval moves closer 
to the dipole axis phase at $\phi=0$. 

In spite of the rather small intrinsic
bifurcation of $2\psi =1^\circ$, the observed bifurcation $\Delta$ can reach 
large values within wide phase intervals that precede and lag the phase of
dipole axis.
If measured from the brightest (main) feature in a pulse profile,
the known double features are located at phase longitudes:
$103.5^\circ$ (B1929$+$10), $51^\circ$, $70^\circ$ (J0437$-$4715),
$-30^\circ$ (B0950$+$08), $-55^\circ$ and $130^\circ$ (J1012$+$5307).
As can be seen in Fig.~\ref{analbeam},  all these numbers, except from 
the last one, lay within the phase interval of strong magnification.
Unfortunately, it is the last location which corresponds to the bright
BEC of J1012$+$5307, the widest-observed so far. However, even this case can
be accounted for by the model, if the dipole inclination
$\alpha$ in J1012$+$5307 is very small or close to $90^\circ$
(in the latter case the phase $\phi=130^\circ$ becomes equivalent
to $\phi=-50^\circ$, and there is no difference between the
main pulse and interpulse).

\subsubsection{Possibility of interpulses}

Regions of \emph{strong} radio emission in pulsar magnetosphere are
definitely more localised than shown in Fig.~\ref{skyd}, because the observed
profiles usually have duty cycles much smaller than $360^\circ$. However,
most of the emission that contains double features is \emph{not} localized
within a narrow range of phase. This is the case of 
B1929$+$10, B0950$+$08, and J0437$-$4715.
Narrow radio pulses become a possibility
when azimuthal limitations for the extent of emission are allowed, 
or when the radial extent of
strong emissivity is not too large, yet noticeable, say
$\Delta r \la 0.1\rlc$. 

It is worth to note that even for
very extended emission regions, interpulses do not have to be always
detectable, because of bulk topological properties of pulsar magnetosphere.
One example of this is the outer gap model, in which `the other pole' emission 
is undetectable by a single observer because it is constrained to regions
above the null-charge surface (Romani \& Yadigaroglu 1995; Wang et al.~2011). 
Another example is the slot gap model (Arons 1983; Muslimov
\& Harding 2003) with the large-scale acceleration possible only on
favourable (poleward) magnetic field lines. Thus, in a large fraction of
existing magnetosphere models the invisibility of the other pole is an
inherent property. It is therefore possible that radially-extended
emission streams are present in many objects, but are not directly manifested
by unavoidable second-pole emission.

\section{Piece-wise fits of physical curves to the double notches}

As can be seen in Fig.~\ref{show3b}, 
the altitude of detectable emission changes gradually 
while the sightline is obliquely traversing through the double pattern.
Therefore, the width and height of the sampled microbeam 
also gradually change with
pulse phase, because they depend on the curvature radius $\rho(r)$ and
$r$-dependent emissivity. Obviously, the same is true for double features in 
the absorption version, ie.~for double notches which may have $r$-dependent 
width and depth. Precise fits to double features then require a full 3D
modelling of the stream-cut geometry, ie.~the $r$-dependence of the depth
of the notches, and the $r$-dependence of $\rho$
need to be assumed in addition to $\alpha$, $\zobs$, and 
the specific functional form of the microbeam.
This kind of calculation is feasible with the code of type
that produced the result shown in Fig.~\ref{skyd}. However, the
frequency-integrated (bolometric) microbeam which is now used by the code, 
would need to be replaced by an
observing-band-integrated beam. And the code would have to be
embedded into a fitting routine with some iterative capability 
of heading towards $\chi^2$ minimum. 
Instead of these numerical developments, we make a direct, but
piece-wise fit to the double notches of B1929$+$10 using 
the bandwidth-integrated beam. More precisely, it is assumed 
that the left half
of the notches on average originates from some unspecified
altitude with one curvature radius $\rho(r_A)$, 
and the other half on average originates from another altitude and 
$\rho(r_B)$.
Since the scale of the fitted beam depends only on the curvature 
radius $\rho$, it is only the values of $\rho_i$ that are provided 
by the fit. Any further association of $\rho_i$ with radial distances
$r_i$ is not unique because of unknown viewing geometry ($\dcut$).
As will be seen below, the limited
quality of the best-available
data on the notches of B1929$+$10, makes this procedure reasonable, because
fractional values of $\chi^2/dof$ can be reached even with this simplified
model.

\subsection{Functional form of the fitted microbeam}

\begin{figure}
   \includegraphics[width=0.48\textwidth]{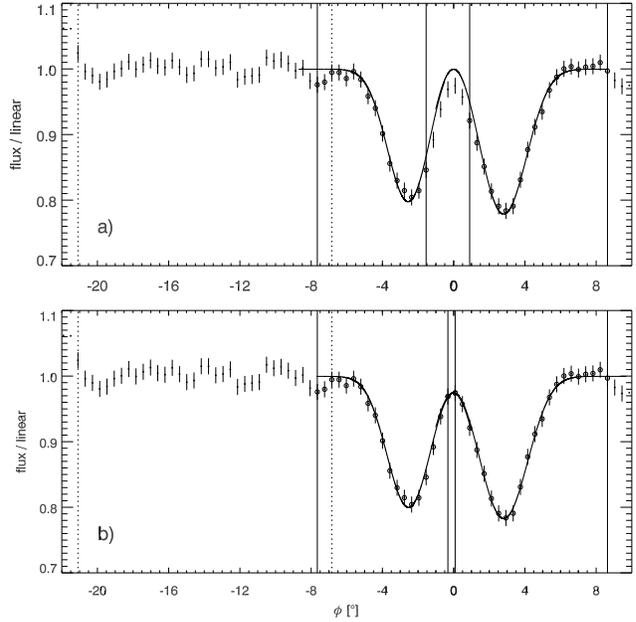}
       \caption{A fit of classical curvature radiation beam (eq.~\ref{int})
   to double notches in the 327-MHz pulse profile of B1929$+$10. Each notch is
fitted separately, within the left-hand side, and right-hand side pair of
solid vertical lines. The pair of dotted verticals marks the phase interval
used to remove the linear increase of flux with phase. In panel a)
the orthogonal polarisation mode is only used
($C_\parallel = 0$). In panel b) we have fixed 
$C_\parallel \simeq 0.13\ C_\perp$ to match the central data points.
The data were collected with the Arecibo Telescope (DRR07);
the bandwidth was $25$ MHz.
       }
      \label{bandfit}
\end{figure}

The radio flux at pulse longitudes surrounding the notches of B1929$+$10
changes linearly (see fig.~1 in DRR07). 
Therefore, before making the fit we remove this
linear trend to get a constant flux normalised to unity (Fig.~\ref{bandfit}).
Then we use the following function to fit the data:
\begin{equation}
I(\phi) = 1 - \int\limits^{\numax}_{\numin} \eta_{\rm crv}\ d\nu,
\label{int}
\end{equation}
where the integration is over the observed frequency bandwidth and
$\eta_{\rm crv}$ represents the shape of curvature radiation microbeam:
\begin{eqnarray}
\eta_{\rm crv} & = & \eta_\parallel + \eta_\perp = \\
&=& \frac{q^2\omega^2}{3\pi^2c}
\left(\frac{\rho}{c}\right)^2
\left[C_\parallel\xi^2 K^2_{\frac{2}{3}}(y) \ +\right.\\
&+& \left.C_\perp\xi\ K^2_{\frac{1}{3}}(y)\sin^2\psi\right],
\label{crv}
\end{eqnarray}
where
\begin{equation}
\xi = 1/\gamma^2 + \psi^2,\ \ {\rm and}
\label{ksi}
\end{equation}
\begin{equation}
y = \frac{\omega\rho}{3c}\xi^{3/2}
\label{yy}
\end{equation}
(eg.~Jackson 1975).
The symbol $\rho$ denotes the radius of curvature of electron trajectory
in the IOF,
$\omega = 2\pi\nu$, $c$ is the speed of light, and $K$'s are the modified
Bessel functions. 
The angle $\psi$ is the azimuth of the line of sight measured
in a frame with z-axis along the instantaneous acceleration vector
(it is the azimuth of sightline measured around the vector
of curvature radius).\footnote{Let us introduce a plane that is orthogonal
to the plane of electron trajectory, and that includes the instantaneous
velocity vector $\vec v$. Let us project the line of sight onto this 
orthogonal plane. $\psi$ is the angle betweeen the projection 
and the plane of electron trajectory. Note that $\psi$ is NOT the angle 
between
the trajectory plane and the sightline itself, as was wrongly 
declared in DRD10.
}
The two terms in the square bracket represent
decomposition of the beam into two parts: one polarised parallel, and 
the other polarised orthogonally
to the plane of electron trajectory. In the absence of the parallel mode
the normalisation constant $C_\parallel$ is equal to
zero, and the microbeam has the double-lobed shape with zero flux in the
centre. Therefore, if spatial convolution effects are not included in the
fit, the central maximum of the double notches, which does 
not reach the `off-notch' level,
is not well reproduced (Fig.~\ref{bandfit}a, with $\chin = 1.1$ for 
the left notch,
and $\chin = 0.6$ for the right notch).
The decreased flux at the central maximum 
probably results from some convolution effects, such as the non-zero
width of the non-emitting/eclipsing region. 
To avoid the complexity of such convolution
(eg.~the selection of transverse plasma density profile),
we instead lower down the maximum by adding a bit of the parallel
mode ($C_\parallel \simeq 0.13C_\perp$, Fig.~\ref{bandfit}b), 
which has a single-peaked pencil-like shape.
Note that the data may well contain no parallel mode at all.
We simply use the non-zero value of $C_\parallel$ as a convenient way
to improve the fit. The spatial convolution seems to be the most natural
explanation for the low flux at the center of `W'.
In the fitting procedure we simply set $\psi = \phi-\phi_c$, 
where $\phi_c$ is the phase of the centre of the `W'.
In the low-frequency limit of $\nu \ll \nucr$,
the shape and size of curvature radiation beam
do not depend on the Lorentz factor $\gamma$ of emitting particles.
Since the frequency of observation shown in Fig.~\ref{bandfit}
is relatively low ($\nu = 327$ MHz), it is reasonable to 
perform the fitting without any reference to $\gamma$.
To minimize the number of fit parameters in this way,
we simply set $\gamma$ to some `very high' value (eg.~$10^7$) to
ensure that $\nu \ll \nucr(\gamma, \rho)$.  
The width scale
of the beam is then changed only through changes of $\rho$. 
Therefore, the fitted
value of %$\rho$ 
curvature radius also incorporates any influence of viewing
geometry on the scale of the notches. Thus, the fit actually provides us 
with the value of $\rho_{\rm min} = \rho\sin^3\zeta\sin^3\dcut$, where the 
subscript `min' has the meaning of minimum value of $\rho$ for 
both $\zeta$ and $\dcut$ equal to $90^\circ$.
In spite of the wide bandwidth ($\Delta\nu/\nu = 8\%$), the frequency 
integration in eq.~\mref{int} does not change the beam shape considerably.
The bandwidth-integrated beam stays similar 
to the frequency-resolved shape of eq.~\mref{crv}.

As can be seen in Fig.~\ref{bandfit}, the curvature radiation beam
reproduces the notches of B1929$+$10 very well. Even when no account is done
for the lowered flux of the central minimum (Fig.~\ref{bandfit}a), 
the shape of the curvature subbeam neatly reflects the steepness of the
outer walls of the `W', as well as the bulk width of the minima, as compared
to the separation between them. With the central maximum lowered down for
the price of one additional fit parameter ($C_\parallel$), the quality of the fit
reaches $\chin = 0.7$ and $0.2$ for the left and right-hand side notch
respectively (Fig.~\ref{bandfit}b). 

\subsubsection{Curvature radii}

The fitted values of curvature radius are: $\rho_{\rm min} = 1.17$, and 
$0.85 \cdot 10^5$ cm
for the leading and trailing notch, respectively. These values assume 
orthogonal and equatorial cut through the stream ($\dcut = \zeta = 90^\circ$)
and should be understood as the lower limits for $\rho = \rho_{\rm
min}/(\sin^3\zeta\sin^3\dcut)$. 
The value of $\rho_{\rm min}$ fitted for the leading notch is somewhat larger than 
for the trailing one. %Such a small difference in $\rho$ 
This seems to be 
natural for the double notches of B1929$+$10, because
small $\dcut$ is likely for this object: the interpulse suggests
that $\alpha\sim\zeta \sim90^\circ$, and 
the notches are located $103.5^\circ$ after the main pulse, ie.~within the
region of strong amplification through the small-$\dcut$ effect 
(Fig.~\ref{analbeam}). Note that already for moderately small 
$\dcut  = 28^\circ$
(and $\zeta=90^\circ$), the inferred value of $\rho$ reaches $10^6$ cm,
whereas for $\dcut=13^\circ$, $\rho = 10^7$ cm. The fitted size of the double
notches ($\Delta = 5.37^\circ$ at $\nu=327$ MHz) 
is then compatible with \emph{dipolar} values of near-surface $\rho$
in the closed field line region.

In the plane of the magnetic equator the curvature radius of dipolar 
B-field lines is equal to:
\begin{equation}
 \rho_{eq} \equiv \rho(\theta_m= 90^\circ) = \frac{r}{3},  
\end{equation}
where $\theta_m$ is the magnetic colatitude of the emission point,
measured from the dipole axis. Thus, for every pulsar with radius 
$R_{ns} = 10^6$ cm, the minimum curvature radius available in its dipolar
magnetosphere is equal to $\rho = 3.3\cdot 10^5$ cm.

In B1929$+$10 the double notches
are observed at $\phi\simeq103.5^\circ$, ie.~roughly at the right angle
with respect to the dipole axis. Therefore, 
it is useful to find the surface
value of $\rho$ at the position where the B-field is pointing 
at the right angle
$\theta_B=90^\circ$ with respect to the dipole axis. The magnetic colatitude
for this point is equal to $\theta_m \simeq 54.73^\circ$
(more precisely: $\sin^2\theta_m = 2/3$), and the corresponding curvature
radius is:
\begin{equation}
\rho_\perp \equiv \rho(\theta_B = 90^\circ) = \frac{\sqrt{3}}{2}\thinspace r,
\end{equation} 
which at the star surface gives $\rho = 0.866 R_{ns}\simeq 10^6$ cm.
Thus, even in the presence of only modest geometrical amplification
($\dcut \simeq 30^\circ$, $\zeta \simeq 90^\circ$) the dipolar curvature 
radii in the
closed field line region are consistent with  
the observed width of double notches in
B1929$+$10.
%
%
% the equator
% between the rim of the polar cap and the
%equator, $\rho$ decreases from $4.3\cdot 10^7$ cm to $3.3\cdot 10^5$ cm.
%The equatorial value of $\rho = 3.3\cdot 10^5$ cm is common 
%for any pulsar with radius $R_{ns} = 10^6$ cm. It is the smallest 
%\emph{dipolar}
%curvature radius available in the magnetosphere of every pulsar.

%However, if one assumes
%that $\dcut$ is not far from $90^\circ$, then 
%
In the unlikely case of large $\dcut$, the substellar values of
$\rho_{\rm min}$ would have to 
be interpreted
in terms of small-scale distortions of dipolar B-field
(Harding \& Muslimov 2011; Ruderman 1991).
Another option,
at least in principle, would be the gyration radius 
of primary particles near the light cylinder:
\begin{equation}
r_B = 1702\ {\rm cm}\ \frac{m}{m_e}\frac{\gamma}{B},
\label{gyrad}
\end{equation}
where $m/m_e\equiv m^\prime$ is the mass of radiating charges in units 
of the electron mass
$m_e$, and $B$ is in Gauss.
This radius translates into the bifurcation angle of:
\begin{equation}
2\psi = 14.4^\circ \left(\frac{B{\rm[G]}}{m^\prime\gamma\nu_9}\right)^{1/3},
\label{psilc}
\end{equation}
which is broadly consistent with observations for $B\la10^5$ G and
$\gamma\ga10^6$. However, to produce strong and unsmeared
double features, the gyrational motion
would have to be spatially ordered within a large volume 
of magnetosphere. This does not immediately seem to be natural.

\subsection{Separation as a function of frequency}

\begin{figure}
   \includegraphics[width=0.48\textwidth]{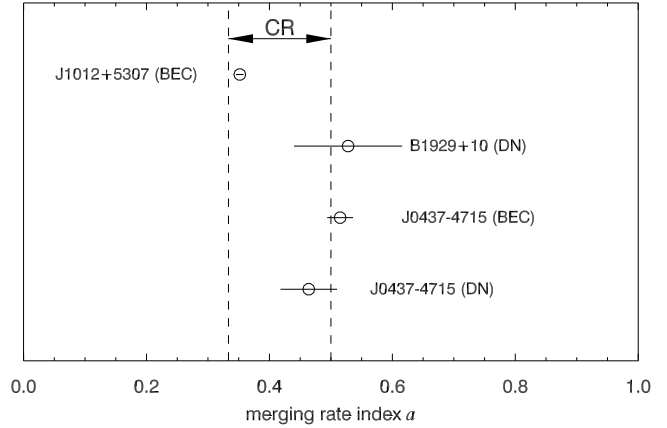}
       \caption{Average merging-rate index $a$ in the relation
$\Delta \propto \nu^{-a}$. 
`BEC' refers to bifurcated emission components,
whereas `DN' to double notches. Dashed verticals mark the range of $a$
expected if the doubleness is caused by the double-lobed microscopic 
beam of curvature radiation (eq.~\ref{int}). The notches of B0950$+$09
(not included in this figure) are also consistent with $a=1/2$ (see fig.~9
in DRR07). 
       }
      \label{mindex}
\end{figure}

In view of the remarkable agreement seen in Fig.~\ref{bandfit}, it is needed
to verify if the separation $\Delta$ between the minima (or maxima)
in double features is also consistent with the properties of curvature
radiation. According to the classical electrodynamics, at low frequencies
($\nu \ll \nucr$) the separation $\Delta$ decreases with increasing $\nu$
according to eq.~\mref{psi}, ie.~according to the power-law $\Delta \propto
\nu^{-a}$ with the index $a = 1/3$. The bifurcation angle in this case
is $\psi \gg 1/\gamma$, where $\gamma$ is the Lorentz factor of the emitting
particles.
At the high-energy end of the emitted
spectrum, ie.~within the high-energy spectral decline,
the features merge faster, with the index $a = 1/2$ (eg.~Jackson 1975),
and the bifurcation becomes very small: $\psi \ll 1/\gamma$.
A simple arithmetic average then allows us to expect an `average' merging 
rate of $a = 0.417$ around the spectral maximum.
It is worth to emphasize that many models of \emph{coherent} 
curvature emission 
follow the same $\psi(\nu)$ as the non-coherent curvature radiation.
Such models include those based on bunching of charges 
(Ruderman \& Sutherland 1975), as well as some
maser mechanisms (Luo \& Melrose 1992; Kaganovich \& Lyubarsky 2010), 
 
The theoretical range of merging rate for the curvature 
radiation
is compared to the available data on double features in Fig.~\ref{mindex},
which includes data on both the emission (BEC) and absorption features (DN).
The figure takes into account the new data on J0437$-$4715 from Yan et
al.~(2011), as well as unpublished low-frequency data on this object.
For this reason the two points for J0437$-$4715 lay closer to $a = 1/2$
than in fig.~6 in DRD10. 

One can see that within the statistical $1\sigma$ errors, 
the measured merging rate
is consistent with the one expected for the curvature beam.
Inspection of $\Delta$ as a function of $\nu$ may even be revealing
some decrease of $a$ at lower frequencies (see fig.~6a in DRR07).
We remind here that it is not the merging rate that differentiates
the curvature radiation from the other emission processes, because the latter
can also have $a \la 1/2$ (eg.~Melrose 1978; DRD10). However, if the origin of the
observed doubleness is not related to the microscopic beam, 
this would possibly be manifested by merging index values from
outside of the range $\left[1/3, 1/2\right]$. The large number of indices near $a =
1/2$ may be considered to be not quite natural for the curvature radiation, 
because
double features should be more easily observable at low $\nu$, where
the beam is wider, and the merging index $a$ is closer to $1/3$. However, 
double features at high $\nu$ can be more easily affected by blurring 
effects, which can also bring the maxima increasingly closer together,
thus increasing $a$. We therefore conclude that the analysis of merging rate
is broadly consistent with the curvature origin of double features.

\subsection{Fitting the BEC of J1012$+$5307}

We have also made several fits to the BEC of J1012$+$5307, in which the
outer wings are much less steep than the outer wings of the double notches
in B1929$+$10. Therefore, we have used functions that are
a convolution of some bell-shaped plasma-density profile (like the Gauss or
Lorentz functions)
with the microbeam of eq.~\mref{int}. This type of model has appeared to be 
unable to reproduce the BEC: before the flux in the wide 
outer wings has been reached, the flux in the central minimum was already too
large. We conclude that with the use of the microbeam from eq.~\mref{int},
the shape of the outer wings of the BEC in J1012$+$5307 cannot be reproduced
by convolving it with bell-shaped density functions. 
Another convolution that is possible for the simple case of $\nu \ll \nucr$, 
and which remains to be tried out,
is a spread in curvature radii $\rho$, which can act
similarly to the integration over a wide frequency range (much wider than the
bandwidth).
Yet another possibility 
is the case of $\nu \simeq \nucr$, ie.~the observation frequency being
close to the characteristic frequency of curvature spectrum, in which case
the opening angle is $\psi \sim 1/\gamma$, where $\gamma$ is the electron
energy in units of $mc^2$. Finding the shape of BEC in such a case would 
require us to make integration over the electron energy distribution
(in addition to the spatial convolution).

\section{Conclusions}

We conclude that the stream-cut geometry is a successful model for such
components in radio pulse profiles like the \emph{apparently} `conal'
components in J0437$-$4715, or the BEC in J1012$+$5307. Such geometry
naturally explains the fixed, frequency-independent location of these
features in the profiles. The components stay at a fixed phase not because
of the `lack of RFM', but because the emission region is narrow in azimuth
(stream-like emitting region, or spoke-like, if there are more streams).
 If the RFM is added to this picture, 
it does not move
the components away from the profile center. Instead, in the case of
non-orthogonal cut through the stream, the RFM is making the double features
asymmetric at all frequencies, except from at $\nusym$. This conjecture
is valid for stream-like emitters with RFM regardless of the actual
origin of the doubleness, ie.~regardless of whether the bifurcation
has the micro- or macroscopic origin.

If we assume that the doubleness reflects the shape of the microphysical
beam, then the curvature radiation in the extraordinary mode
becomes a natural process, because: 1. it is intrinsically double-lobed 
(bifurcated), which is a necessary condition for producing \emph{deep}
double notches (DRD10); 2. the extraordinary part of the curvature beam
reproduces the shape of notches in B1929$+$10; 
3. the rate of merging of double features
with increasing frequency is consistent with the curvature origin;
4. the width-scale of double features in B1929$+$10 and J0437$-$4715 is
consistent with dipolar curvature radii;
5. the high polarisation of double features is consistent with the
single-mode nature of the double-lobed beam.

This said, it is not quite clear why some of the observed double features
have the width of $\ga 7^\circ$ at 1 GHz, despite the intrinsic beam
is expected to be only $1^\circ$ wide. However, the quite large  probability 
of strong
geometric magnification, along with the possibility of smaller-than-dipolar
$\rho$, do not really allow us to consider this issue as 
a serious problem for the curvature microbeam model.
As recently noted by Yan et al.~(2011), who verify fairly large sample 
of high-quality 
profiles of MSPs, bifurcated features are rare among MSPs.
This is in line with the special requirements that need to be fulfilled
for these features to be observable. The requirements apparently 
include at least the following: 
1. the spatial extent (azimuthal width of the
stream) must not be too large to blur the feature; 2. the feature be best
magnified by geometrical effects to be wide and easily detectable.
It is not clear why we do not observe features with $\Delta \la
1^\circ$, that fulfill the condition no.~1, but are not magnified
geometrically. Generally, however, the intrinsic
narrowness of the curvature beam seems to be consistent with the rarity
of double features in the observed profiles. This is because
for even a small azimuthal extent of emission region, the doubleness should
become hidden by spatial convolution effects.
  
We conclude that the asymmetry of double features at different frequencies, 
the shape of notches in B1920$+$10, the way they merge, their high
polarisation, and their fixed phase locations
can all be understood in terms of sightline cuts through streams
of plasma that emits curvature-radiation in the orthogonal polarisation 
mode.

%koniec
\section*{acknowledgements}

We are grateful to Joanna M. Rankin for providing
us with Arecibo data on B1929$+$10.
%JD is indebted to K.~Lazaridis, A.~Jessner,
%R.~Manchester, and J.M.~Rankin
%for providing us with pulsar data. 
This work was supported by the grant N203 387737
of the Polish Ministry of Science and Higher Education.
%and the Polish Astroparticle Network 621/E-78/SN-0068/2007.

\end{document}